# AC susceptibility study of superconducting $YBa_2Cu_3O_7$:$Ag_x$ bulk composites (x = 0.0-0.20): The role of intra and inter granular coupling

Poonam Rani, Rajveer Jha and V.P.S Awana[*]

Quantum Phenomena and Application Division, National Physical Laboratory (CSIR)
Dr. K.S Krishnan Marg, New Delhi-110012, India

We report the effect of silver addition on superconducting performance of bulk YBCO ($YBa_2Cu_3O_7$) superconductor. All the studied samples are prepared by conventional solid-state reaction method. Rietveld fitted X-ray diffraction data confirmed the single phase formation for all the studied samples. Detailed AC susceptibility measurements as a function of driven AC amplitude (1Oe-17Oe) of these samples revealed the enhancement of grains coupling with increasing Ag content in YBCO+$Ag_x$ composite system. 10wt% Ag added YBCO superconductors exhibited the optimum inter granular coupling. The Scanning Electron Microscopy (SEM) observations indicate an increase in the grains connectivity in terms of narrow grain boundaries for doped samples. The average grain size is found to increase with Ag doping. It is concluded that limited addition of Ag in bulk YBCO superconductor significantly improves the grains coupling and as result optimum superconducting performance. YBCO+Ag composites could prove to be potential candidates for bulk superconducting applications of the studied high $T_c$ system.

*Corresponding author: e-mail – awana@mail.npindia.org
Tel.: +91 11 45609357; Fax; +91 11 45609310.

# Introduction

Since the discovery of $YBa_2Cu_3O_7$ (YBCO) compound [1], various elements are added in YBCO system for increasing its critical temperature ($T_c$) and critical current density ($J_c$) [2-4]. In various bulk copper oxides superconductors including YBCO, the grain boundaries play crucial role in deciding their superconducting performance. High critical current density ($J_c$) superconductors are useful in the field of electric power applications. Polycrystalline high $T_c$ superconductor materials can be described as arrays of superconducting grains being weakly coupled by Josephson junctions. The $J_c$ values of these superconductors are limited by these weakly coupled grains [2-5]. Some of the possible reasons for the formation of these weak links are mis-orientation of grain boundaries and composition variations at the grain boundaries [3-4]. The low value of the grain boundary critical current density in polycrystalline samples is a known problem for large-current applications [1-4]. It can be improved through grain alignment, grain-boundary doping and optimization of the microstructure to minimize the effective grain-boundary area [5]. One of the effective ways to minimize the grain boundary area is to increase the size of superconducting grains.

Addition of Ag in various high $T_c$ cuprate oxide superconductors (HTSc) has resulted in improving the superconducting performance of these systems [6-10]. In this shot note, we present the results of XRD, microstructure (SEM) and detailed AC susceptibility on pure and Ag added YBCO superconducting samples. It is found that limited Ag addition in YBCO superconductor results in improved superconducting properties of bulk samples, as revealed from detailed AC susceptibility results. The SEM results indicated that with Ag addition the effective superconducting grains size is improved and at higher Ag content the same is seen as dispersed at inter-granular spaces. There is hardly any change in the crystallographic parameters with Ag addition. It seems Ag is not substituted in the lattice but rather has improved the grain size by seemingly decreasing the phase formation temperature for low content (10wt% Ag) and at higher contents Ag is mostly dispersed at grain boundaries. It is thus concluded that limited addition of Ag in YBCO superconductor improves its superconducting performance by increasing the superconducting grains size.

## Experimental

The samples of YBa$_2$Cu$_3$O$_{7-\delta}$+Ag$_x$ series are prepared through conventional solid state reaction route with nominal composition x= 0, 0.05, 0.10, 0.15 and 0.20wt%. High purity powders of Y$_2$O$_3$, Ag$_2$O, BaCO$_3$ and CuO and (99.99%) with exact stoichiometric ratio are mixed. After initial grinding, calcinations are done at 870$^o$C, 890$^o$C and 910$^o$C temperatures with intermediate grinding. In each calcinations cycle cooling is done slowly (2K/minute) and samples are re-ground well before the next cycle. After final calcination the samples are pressed into rectangular pellet form and sintered at 920$^o$C for 48 h in air. Finally the samples are annealed with flowing oxygen at 920$^o$C for 12 h, 600$^o$C for 12 h and 450$^o$C for 12 h and subsequently cooled slowly to room temperature. All the samples are characterized by the X-ray powder diffraction technique using Rigaku X-ray diffractometer (Cu-K$_\alpha$ line). Rietveld analysis of all samples is performed using Fullprof program. Preliminary DC and detailed susceptibility measurements are done on Physical Properties Measurement System (Quantum Design-USA PPMS-14T). The SEM images of the samples are taken using *ZEISS* EVO MA-10 Scanning Electron Microscope.

## Results and Discussion

Fig 1 shows the Rietveld fitted *XRD* patterns of the studied YBa$_2$Cu$_3$O$_7$+Ag$_x$ samples. All the samples are crystallized in single phase having orthorhombic structure within *Pmmm* space group. Some small intensity peaks of Ag are also observed in Ag doped YBCO samples, in particular for higher Ag content samples. The lattice parameters of all the studied samples are given in Table 1. It can be seen from table 1, that the orthorhombicity of Ag doped YBCO is mostly unchanged with Ag addition. This is in agreement with earlier reports [6,7]. Our XRD results undoubtedly prove that the studied YBCO+Ag$_x$ samples are of reasonably good quality.

Detailed AC susceptibility for the pure YBCO sample with temperature in superconducting regime (2-100K) is shown in Figures 2(a) and 2(b). Figure 2(a) exhibits the real part of AC susceptibility ($\chi'$) at different (1Oe-17Oe) ac magnetic field amplitudes. It is important to note that the DC bias field is zero. It is not easy to get zero DC field in particular on a high field (14 Tesla) PPMS. One need to ramp the DC field in various + and − field situation to reach exact zero DC bias field [11]. The AC susceptibility measurements sequence is started only after achieving zero DC bias field above superconducting transition temperature ($T_c$) say at 100K or above in present case. Now we discuss the detailed real part AC susceptibility ($\chi'$) of pure YBCO being done at various (1Oe-17Oe) amplitudes and zero DC bias field. The diamagnetic transition is seen below 90K with a broad step below say 86.5K at 1Oe ac magnetic field amplitude. Clearly the real part of the AC susceptibility consists of two transitions, corresponding to the flux removal from the intra-grain (90K) and inter-grain (86.5K) regions of the YBCO superconductor. This behavior is known for the HTSc cuprate superconductors, due to their strong inter granular nature and the near insulating grain boundaries existing between superconducting grains [2-6]. With an increase in AC magnetic field amplitude from 1Oe to 17Oe, interestingly though the diamagnetic transition starts from same onset temperature of 90K (intra grain $T_c$), the broad step like transition is shifted from 88.5K to 68.5K (inter grain $T_c$). This is clear indication of the strong granular nature of the studied polycrystalline bulk YBCO superconductor. The inter granular weak links in the studied YBCO superconducting system are coupled through the near insulating grain boundaries and hence the flux exclusion from coupled SIS (superconductor-insulator-superconductor) junction is strongly dependent on the AC amplitude. This is the reason that the inter grain $T_c$ of 88.5K is shifted to 68.5K with an increase in AC amplitude from 1Oe to 17Oe. The relative shift in inter grain $T_c$ of a granular superconductor defines the effective coupling of to superconducting grains. In any granular superconductor, better is the superconductor grains coupling, the smaller is relative shift in inter grain $T_c$ with AC amplitude.

The imaginary part of the AC susceptibility ($\chi''$) for pure YBCO sample is depicted in Figure 2(b). Clearly two peaks are observed in the $\chi''$ versus T plots being taken at various AC amplitudes from 1Oe to 17Oe. The first peak is relatively lower in size than the second one. The first peak occurs at around 89K and second at around 86.28K with AC amplitude of 1Oe. The two peaks correspond to the intra and inter granular $T_c$ of the YBCO, similar to that as in case of $\chi'$ versus T (Fig.2a). With an increase in AC amplitude from 1 Oe to 17Oe, the first peak though remains unaltered at around 89K, the second peak is shifted from 86.28K (1Oe) to 65.88K (17Oe). This clearly indicates that though intra grain $T_c$ is unaltered with AC amplitude, the intra grain $T_c$ is decreased significantly. The second peak temperature being defined as ($T_p$) is given in Table 2 with various (1Oe-17Oe) AC amplitudes for the YBCO sample. In a way, the imaginary part of AC susceptibility is replica of the real part of the same. Because relative shift of $T_p$ with AC amplitude magnetic field (H) i.e. $dT_P/dH$ is an important parameter, which indirectly tells about the coupling strength of the superconducting grains in a granular SIS/SNS system. In case of YBCO the $dT_P/dH$ comes out to be (86.28K-65.88K)/(17Oe-1Oe) = 1.30K/Oe. Higher is $dT_p/dH$ the weaker is the inter grain coupling in a granular bulk superconductor.

The real part AC susceptibility of YBCO+10wt%Ag sample at various amplitudes (1Oe-17Oe) is shown in Fig. 3(a). Similar to that as in case for YBCO in Fig. 2(a) two transitions are seen in superconducting region i.e. below 90K. The first one being intra grain and the second one is due to intra grain superconducting transition. Also, though the intra grain transitions remains unchanged at around 90K the inter grain one shifts to lower temperature with increase in AC amplitude from 1 Oe to 17Oe. Similar trend is seen in imaginary part AC susceptibility (Fig. 3b) of the YBCO+10wt%Ag sample. Although qualitatively the AC susceptibility results of pure YBCO and YBCO+10wt%Ag sample look similar to each other, quantitatively the same are very different from each other. This is clear from Table 2, where $T_p$ is tabulated against the AC amplitude for both pure and 10wt% Ag doped samples. For example at 1Oe the $T_p$ of pure YBCO is 86.28K, which is increased to 87.99K for 10wt% Ag doped sample. An increase in $T_p$ for 10wt%Ag sample in comparison to pure YBCO is clear indication of an increased inter gain coupling in the Ag doped sample. Both inter and intra grain superconducting transitions approaches to each other for a better coupled superconductor with least grain boundaries. It seems the inter grain coupling in 10wt%Ag sample is improved in comparison to the pure YBCO. This is clear, when one examines the $dT_P/dH$ for 10wt%Ag sample. In case of YBCO+10wt%Ag sample, the $dT_P/dH$ comes out to be (87.99K-71.51K)/(17Oe-1Oe) = 1.02K/Oe. Interestingly the $dT_p/dH$ for pure YBCO is 1.30K/Oe, which is 30% more than the one (1.02) for YBCO+10wt%Ag sample. As mentioned earlier Higher is $dT_p/dH$, the weaker is the inter grain coupling in a granular bulk superconductor. This is clear from results in Figures 2 and Figure 3 that the inter granular coupling between superconducting grains in YBCO+10wt%Ag sample is stronger than as in pure YBCO.

The results of AC susceptibility for higher Ag content i.e., YBCO+15, 20wt%Ag samples are shown in Figures 4(a, b) and 5(a, b). For both 15 and 20wt% Ag samples, the first peak of imaginary AC susceptibility is quite prominent and remains unchanged at its position, but the second peak position is shifted relatively faster with increase in AC amplitude. In fact the peak is not complete for higher amplitudes of above say 11 Oe for 20wt%Ag sample. Qualitatively the peak shift of these two samples can be seen from table 2, where $T_p$ values at various amplitudes are given. Let us mention that the AC susceptibility measurements on these samples are done in identical situations, strictly with zero DC field. It seems that with addition of silver to YBCO superconductor beyond 10wt%, the inter grains coupling is not improved. Although the intra grain transition is strong for higher Ag content samples the inter grain coupling is sufficiently weakened with $dT_P/dH$ of above 2.0K/Oe for 15wt% Ag sample and above 3.7K/Oe for 20wt% Ag sample. It seems the excess Ag in case of 15wt% and 20wt% Ag samples lies at grain boundaries in bulk chunks instead of its fine distribution as in case of 10wt% Ag sample. The inter grains coupling could be improved two way i.e., either by increase in grain size and effective decrease in grain boundaries or by fine distribution of metallic Ag at grain boundaries. This is possibly the case for 10wt% Ag sample. To answer the possibility, we took SEM micrographs of the studied samples and results are shown in Figure 6(a-d)

Fig 6 (a-d) shows the SEM images of the pure and Ag doped YBCO samples. It is observed from the SEM images that the pristine sample is having though better surface texture, but high porosity and lower density. The grain connectivity is increased with Ag doping. The circular grains are seen pure sample, while the doped samples contain the elongated grains. Interestingly the grain size has initially increased for 10wt% Ag sample. For Ag content samples larger chunks of Ag are also seen with the YBCO grains.

In conclusion, detailed AC susceptibility studies are carried out for pure and Ag added YBCO bulk polycrystalline superconductor. It is found the inter grain coupling is sufficiently improved for Ag added samples till 10wt% and is decreased for higher Ag contents.

**Figure Captions**

Figure 1 Fitted and observed room temperature X-ray diffraction of various YBCO+Ag samples

Figure 2 (a) Real part and (b) imaginary part AC susceptibility of pure YBCO sample at various AC amplitudes.

Figure 3 (a) Real part and (b) imaginary part AC susceptibility of YBCO+10wt%Ag sample at various AC amplitudes.

Figure 4 (a) Real part and (b) imaginary part AC susceptibility of YBCO+15wt%Ag sample at various AC amplitudes.

Figure 5 (a) Real part and (b) imaginary part AC susceptibility of YBCO+20wt%Ag sample at various AC amplitudes.

Figure 6 SEM (Scanning electron micrographs) of (a) pure YBCO, (b) YBCO+10wt%Ag, (c) YBCO+15wt%Ag and (d) YBCO+20wt%Ag samples.

**Table 1** Lattice parameters, volume, orthorhombicity and quality of fit parameter ($\chi^2$) for various YBCO+Ag samples.

| Compound | a (Å) | b (Å) | c (Å) | Vol. (Å$^3$) | $\chi^2$ | {(b-a)/b}*100 |
|---|---|---|---|---|---|---|
| YBCO | 3.821(1) | 3.886(3) | 11.671(1) | 173.33 | 7.24 | 1.692 |
| YBCO + Ag 10% | 3.819(4) | 3.885(1) | 11.678(1) | 173.32 | 7.38 | 1.687 |
| YBCO + Ag 15% | 3.816(5) | 3.883(2) | 11.677(3) | 173.04 | 6.33 | 1.705 |
| YBCO + Ag 20% | 3.821(3) | 3.886(2) | 11.681(2) | 173.45 | 9.55 | 1.711 |

**Table 2** AC amplitude (1-17Oe) and $T_p$ for studied YBCO+Ag samples.

| AC Amplitude (Oe) | $T_p$ - YBCO K | $T_p$ - YBCO + 10wt% Ag K | $T_p$ - YBCO + 15wt% Ag K | $T_p$ - YBCO + 20wt% Ag K |
|---|---|---|---|---|
| 1 | 86.28 | 87.99 | 78.84 | 70.95 |
| 3 | 82.92 | 86.00 | 70.79 | 59.41 |
| 5 | 80.52 | 84.19 | 65.44 | 48.69 |
| 7 | 78.40 | 82.06 | 61.09 | 40.39 |
| 9 | 76.86 | 80.24 | 56.73 | 33.47 |
| 11 | 73.91 | 77.99 | 52.77 | 26.30 |
| 13 | 71.23 | 76.30 | 49.40 | 23.05 |
| 15 | 68.69 | 74.04 | 44.05 | 17.15 |
| 17 | 65.88 | 71.51 | 42.49 | 9.68 |

Figure 1

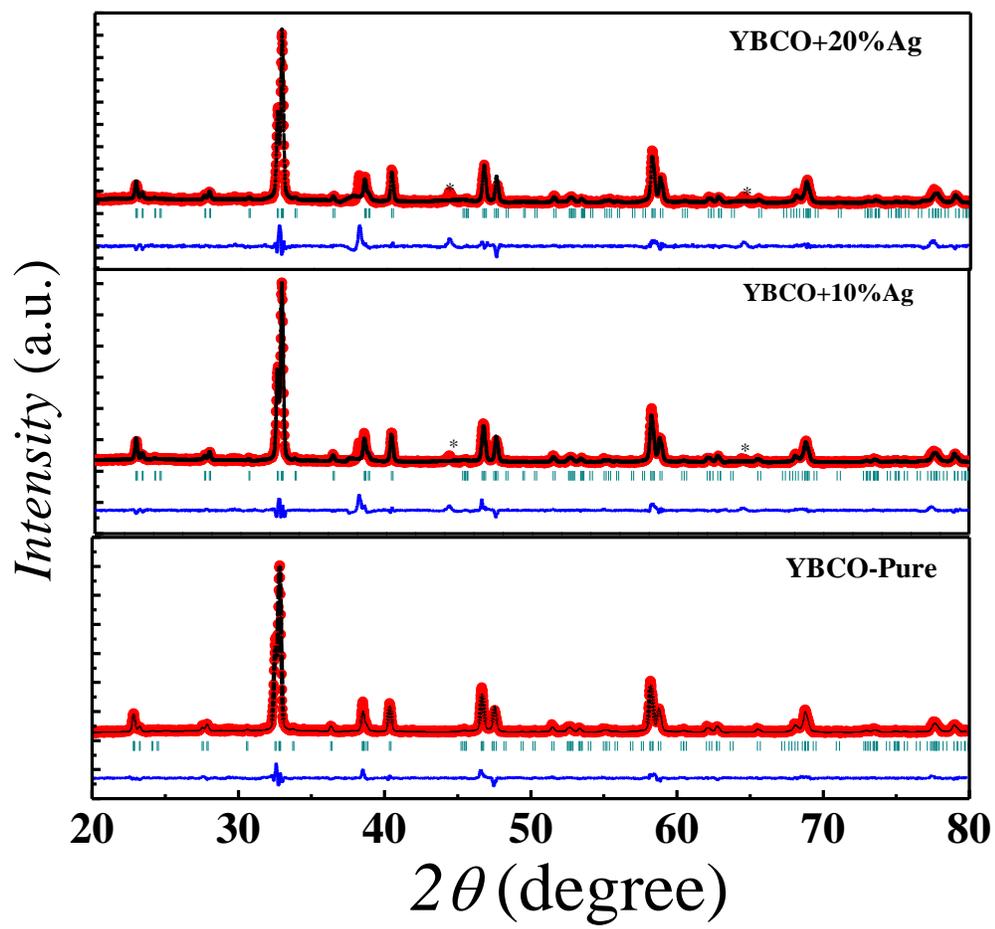

Figure 2 (a)

Figure 2 (b)

Figure 3 (a)

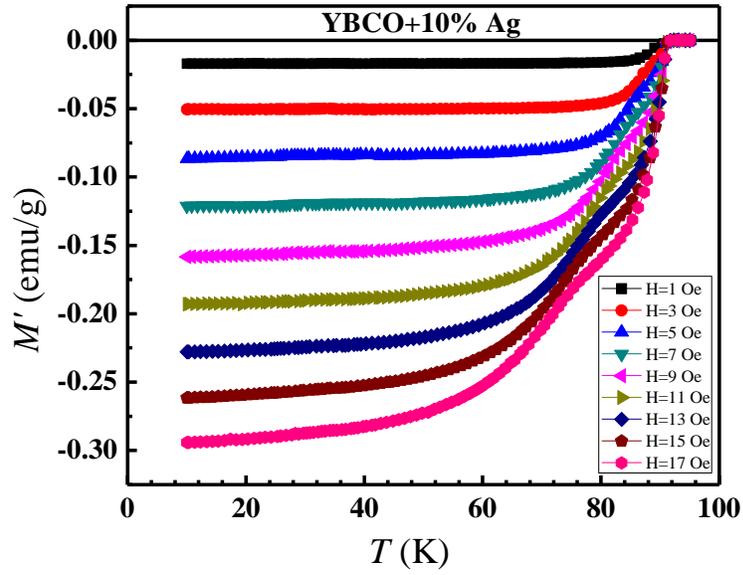

Figure 3 (b)

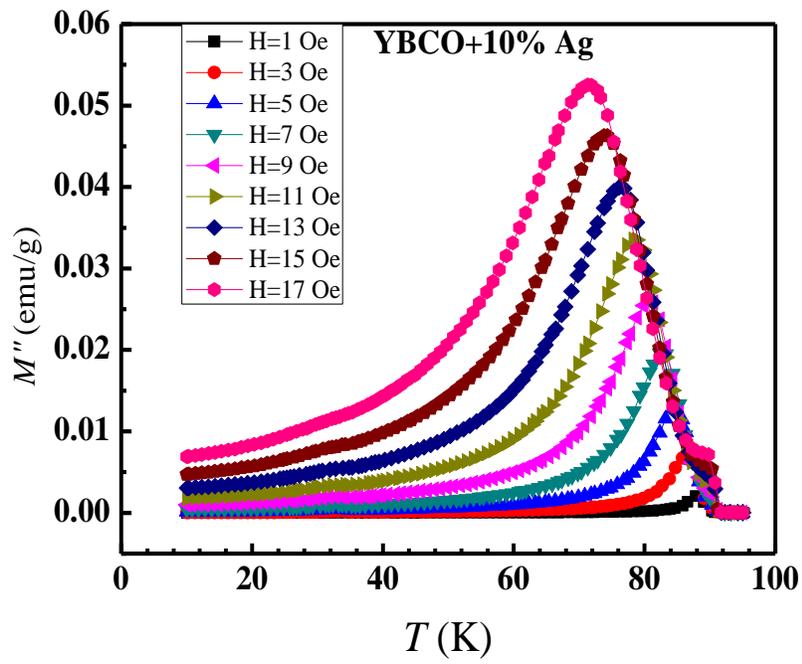

Figure 4 (a)

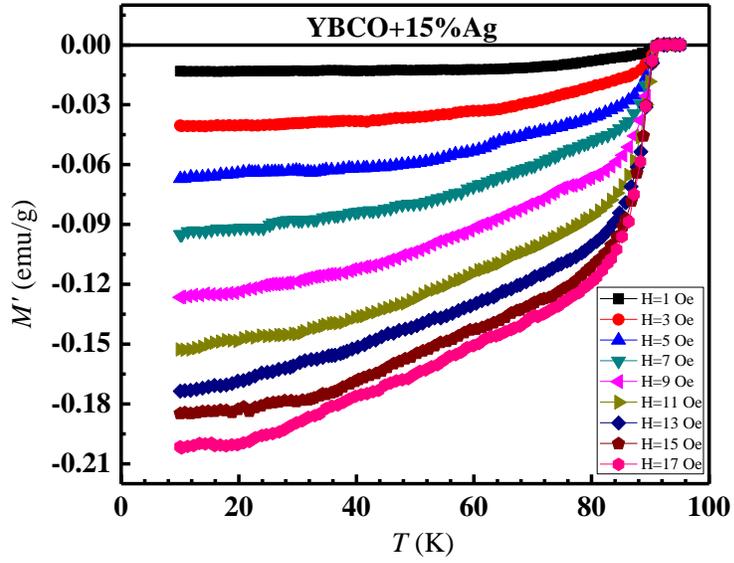

Figure 4 (b)

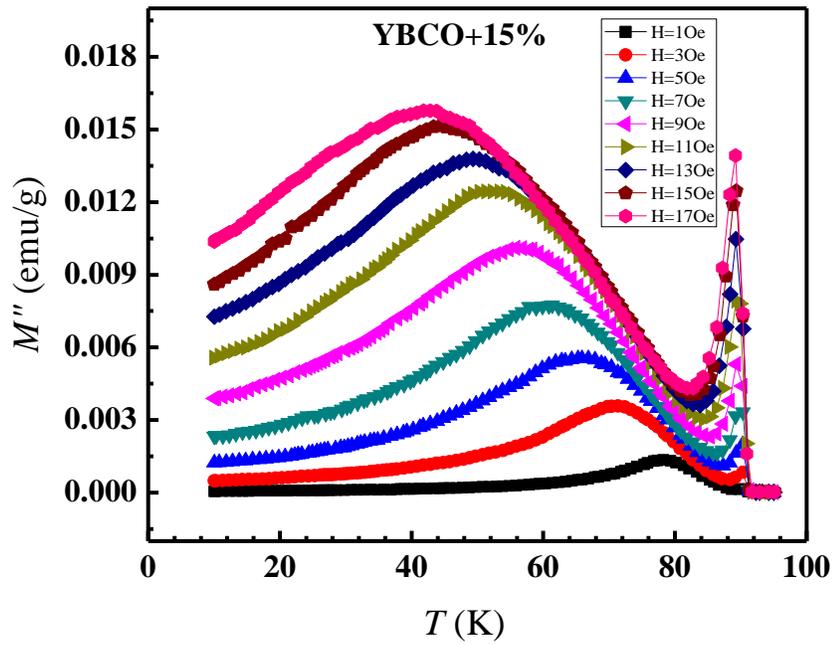

Figure 5 (a)

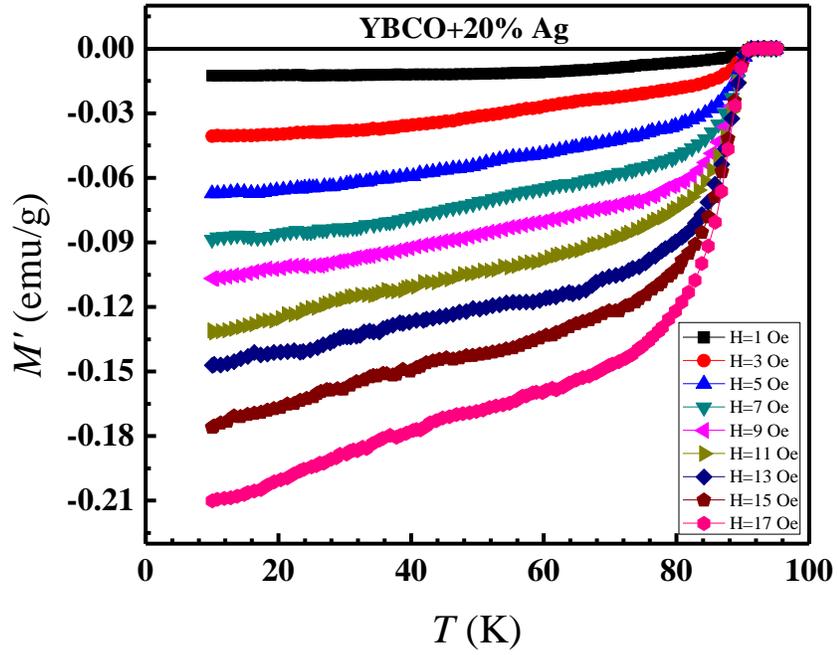

Figure 5 (b)

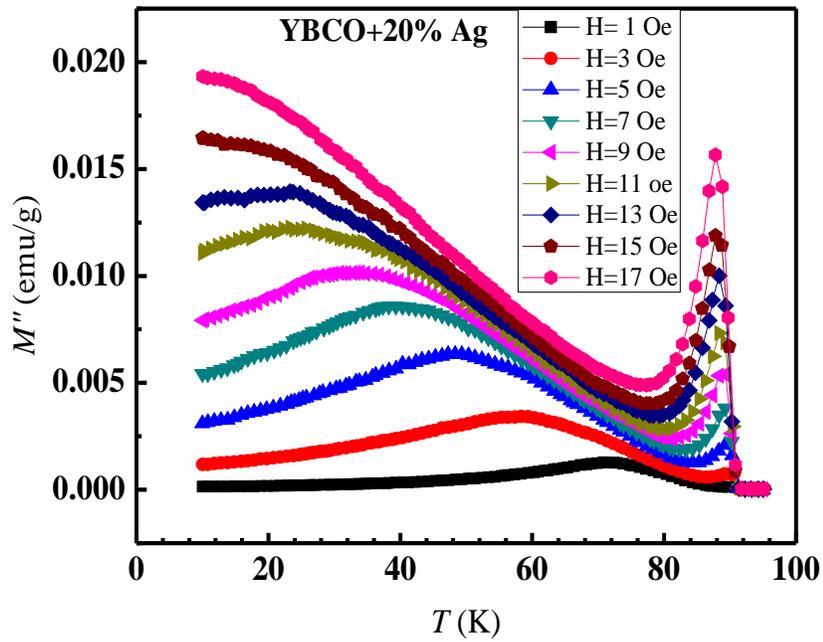

Figure 6

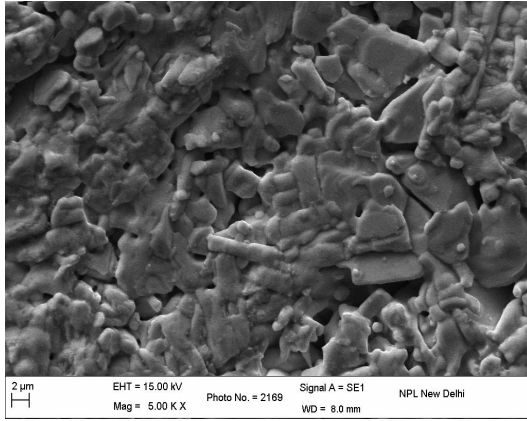
(a)

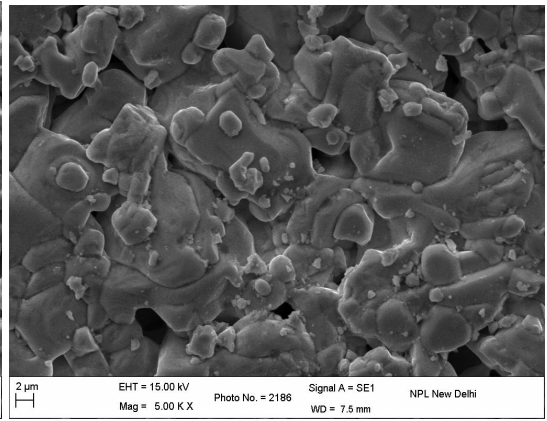
(b)

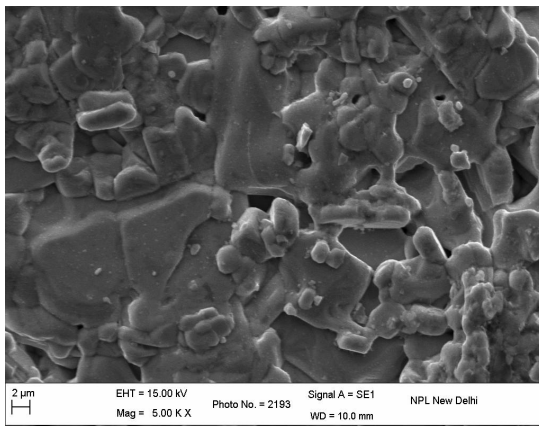
(c)

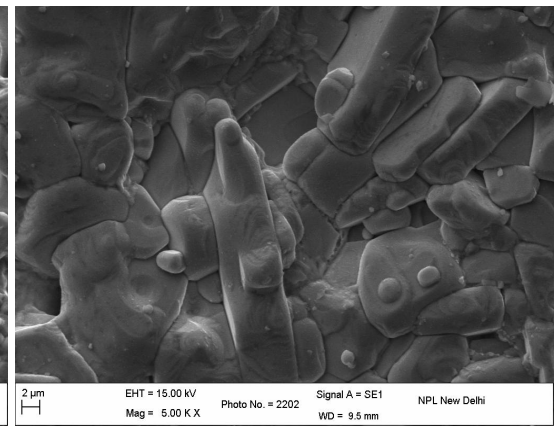
(d)